\documentclass[sigconf,nonacm]{acmart}

\AtBeginDocument{%
  \providecommand\BibTeX{{%
    \normalfont B\kern-0.5em{\scshape i\kern-0.25em b}\kern-0.8em\TeX}}}

\setcopyright{acmcopyright}
\copyrightyear{2024}
\acmYear{2024}
\acmDOI{XXXXXXX.XXXXXXX}

\acmConference[RAID]{Make sure to enter the correct
  conference title from your rights confirmation emai}{September 30 - October 2, 2024}{ Padua, Italy}
\setcopyright{acmlicensed}\acmConference[RAID]{27th International Symposium on Research in Attacks, Intrusions, and Defenses}{September 30 - October 2, 2024}{ Padua, Italy}
\acmBooktitle{RAID, September 30 - October 2, 2024,  Padua, Italy}

\acmPrice{15.00}
\acmISBN{978-1-4503-XXXX-X/18/06}

\usepackage{graphicx} 
\usepackage{algorithmic}
\usepackage{textcomp}
\usepackage{xcolor}
\usepackage{etoolbox}
\usepackage{multirow}
\usepackage{url}
\usepackage{multirow,booktabs}
\usepackage{subcaption}
\usepackage{wrapfig}
\usepackage{float}
\usepackage{float}
\usepackage{tcolorbox}
\usepackage[flushleft]{threeparttable}
\usepackage{relsize}
\usepackage[inline]{enumitem}
\usepackage{amsmath}
\usepackage{color,soul}

\usepackage{algorithm2e}

\newcommand{\revise}[1]{{\color{black}{#1}}}
\newcommand{\note}[1]{{\color{black}{#1}}}

\newcommand{\ourapp}{\textsc{AI2TALE}}
\newcommand{\bx}{{\mathbf{x}}}

\newcommand{\by}{{\mathbf{y}}}

\newcommand{\rqone}{Can our \ourapp~ method solve the phishing attack localization by predicting and giving explanations in a weakly supervised setting? Does the selected information (i.e., sentences) also reflect the human cognitive principles used in phishing emails?}

\newcommand{\rqtwo}{Does the most important (top-$1$) selected sentences from our \ourapp~ method primarily consist of the commonly used and important cognitive principles (i.e., Scarcity, Consistency, and Authority) in phishing emails compared to those of the baselines?}

\newcommand{\rqthree}{Can our proposed information bottleneck theory training term and data-distribution mechanism be successful in improving the data representation learning and supporting the process of learning and figuring out the most important and phishing-relevant sentences in phishing emails?}

\begin{document}

\title{An Innovative Information Theory-based Approach to Tackle and Enhance The Transparency in Phishing Detection}

\author{Van Nguyen}
\affiliation{%
  \institution{Monash University, Australia \\ CSIRO's Data61, Australia}
  \country{}
}
\authornote{Corresponding Author (van.nguyen1@monash.edu). Van Nguyen is a Postdoctoral Research Fellow at the Department of Software Systems and Cybersecurity at Monash University, Australia. Additionally, he is an Affiliate at CSIRO’s Data61, Australia.}

\author{Tingmin Wu}
\affiliation{%
  \institution{CSIRO's Data61, Australia}
  \country{}
}

\author{Xingliang Yuan}
\affiliation{%
  \institution{Monash University, Australia}
  \country{}
}

\author{Marthie Grobler}
\affiliation{%
  \institution{CSIRO's Data61, Australia}
  \country{}
}

\author{Surya Nepal}
\affiliation{%
  \institution{CSIRO’s Data61, Australia}
  \country{}
}

\author{Carsten Rudolph}
\affiliation{%
  \institution{Monash University, Australia}
  \country{}
}

\begin{abstract}
Phishing attacks have become a serious and challenging issue for detection, explanation, and defense. Despite more than a decade of research on phishing, encompassing both technical and non-technical remedies, phishing continues to be a serious problem. Nowadays, AI-based phishing detection stands out as one of the most effective solutions for defending against phishing attacks by providing vulnerability (i.e., phishing or benign) predictions for the data.
However, it lacks explainability in terms of providing comprehensive interpretations for the predictions, such as identifying the specific information that causes the data to be classified as phishing.
To this end, we propose an innovative deep learning-based approach for email (the most common phishing way) \textit{phishing attack localization}. Our method can not only predict the vulnerability of the email data but also automatically learn and figure out the most important and phishing-relevant information (i.e., sentences) in the phishing email data where the selected information indicates useful and concise explanations for the vulnerability. \textit{The rigorous experiments on seven real-world diverse email datasets show the effectiveness and advancement of our proposed method} in selecting crucial information, offering concise explanations (by successfully figuring out the most important and phishing-relevant information) for the vulnerability of the phishing email data. \revise{Particularly, our method achieves a significantly higher performance, ranging from approximately 1.5\% to 3.5\%, compared to state-of-the-art baselines, as measured by the combined average performance of two main metrics Label-Accuracy and Cognitive-True-Positive}.

\vspace{3mm}
\end{abstract}

\settopmatter{printfolios=true}
\maketitle

\section{Introduction}\label{sec:introduction}

Phishing attacks (i.e., \textit{attempts to deceitfully get personal and financial information such as usernames, passwords, and bank accounts through electronic communication with malicious intentions}) have become a serious issue. Nowadays, there are various ways to conduct phishing attacks; the most common method is through the use of emails. Email phishing is crafted to trigger psychological reactions in the users by using persuasion techniques via cognitive principles ~\cite{van2019cognitive} such as scarcity, consistency, and authority.  

According to recent reports, there has been a notable rise in the occurrence of increasingly sophisticated phishing attacks, presenting more formidable challenges for detection and defense. The FBI Internet Crime Report ~\cite{FBIreport} reveals that phishing attacks were the most common cyber-crime type in the United States causing substantial economic losses every year from 2018. Similarly, the Anti-phishing Working Group (APWG) \cite{APWG} reports that the number of phishing attacks has grown by more than 150\% per year since the beginning of 2019. The year 2022 was a record for phishing, with the APWG logging more than 4.7 million attacks. These reports highlight the danger of phishing attacks and the need for a greater understanding of this serious problem.

The widespread adoption of artificial intelligence (i.e., using machine learning-based and deep learning-based approaches) has brought substantial influences and great success in various domain applications such as autonomous driving \cite{chen2023endtoend}, data generations \cite{BERT2018, radford2019language, T52019}, drug discovery \cite{Paul2021}, and malware vulnerability detection \cite{VulDeePecker2018, nguyen2019deep, van-nguyen-dual-dan-2020, ReGVD2021, nguyen2022info, nguyen2022cross, Michael2023VulExplainer}). By leveraging the power of machine learning and deep learning, there have been many efforts proposed for solving phishing attack problems from phishing attack detection \cite{Venkatesh2013, URLNet2018, tyagi2018novel, sahingoz2019machine, yang2019phishing, SOK2019, xiao2020cnn, yang2021phishing} to the investigation of the cognitive bias's impact used in email phishing \cite{phishingstudies2015, van2019cognitive}. It has been proven that using machine learning and deep learning-based algorithms to detect phishing attacks in the early stages is one of the most effective solutions for preventing and reducing the negative effects caused. 
Recently, with ongoing capabilities and improvements in Natural language processing (NLP) and the popularity of large language models, e.g., GPT2 \cite{radford2019language}, BERT \cite{BERT2018}, and T5 \cite{T52019}, some of the early attempts use large language models for phishing attack-related issues have been introduced, e.g., scam-baiting mail servers that can conduct scam-baiting activities automatically \cite{chen2023active}.

Despite more than a decade of research on phishing (i.e., encompassing both technical and non-technical remedies), it still continues to be a serious problem: (i) the problem itself may be intractable, the technical approaches so far may have missed important characteristics (e.g., psychological manipulation techniques used by attackers) of the problem, and (ii) phishing exploits the human as the weakest link so purely technical approaches may not be sufficient \cite{SOK2019}. To overcome these problems, in addition to enhancing human understanding of phishing, phishing attack detection appears as one of the most effective solutions to help human users defend against phishing attacks by providing the vulnerability prediction (i.e., phishing or benign) before they take any actions following the content mentioned in the corresponding data (e.g., the email's content). \textbf{Although phishing detection methods can predict the data's vulnerability label (i.e., phishing or benign), these predictions often lack the explainability to offer meaningful interpretations (i.e., which information causing the data phishing) to the users}. Motivated by this problem, in this paper, we study the following research question:

\vspace{1mm}
\colorbox{gray!10}{
\begin{minipage}{0.92\columnwidth}
"In addition to predicting the vulnerability (i.e., phishing or benign) of the data, how to derive an effective deep learning-based method that can also automatically learn and figure out the most important and phishing-relevant information (i.e., sentences) for providing a useful and concise explanation about the vulnerability of the phishing data to users? (in our paper, we name this problem as phishing attack localization)"
\end{minipage}
}
\vspace{1mm}

To this end, in the scope of our paper, we study the phishing attack localization problem on phishing emails (i.e., the most common phishing way) where we propose an innovative information theory-based model to solve the problem. Our method not only can detect the vulnerability (i.e., phishing or benign) of the email data but also can automatically learn and identify the most important and phishing-relevant information (e.g., sentences) in phishing emails. The selected information helps provide useful and concise explanations about the vulnerability of the phishing email data. \textbf{It is worth noting that the ability to figure out the important and phishing-relevant information that causes the associated email data to be classified as phishing to provide a corresponding comprehensive interpretation is the main difference between phishing attack localization and phishing attack detection}.

\vspace{1mm}
In summary, our key contributions are as follows:
\begin{itemize}
\vspace{0mm}
\item We study an important problem of phishing attack localization aiming to tackle and improve the explainability (transparency) of phishing detection. Automated machine learning and deep learning-based techniques for this problem have not yet been well studied.\vspace{0.5mm}
\item We propose a novel deep learning-based framework derived from an information-theoretic perspective and information bottle-neck theory for phishing attack localization. Our proposed approach can work effectively in a weakly supervised setting (refer to Section \ref{sec:problemstatement} for details), hence providing an important practical solution for defeating phishing attacks.\vspace{0.5mm}
\item Besides proposing a deep learning-based approach for phishing attack localization, based on the explainable
machine learning and email phishing domain knowledge, we propose to use some appropriate measures including \textit{Label-Accuracy}, \textit{Cognitive-True-Positive}, and \textit{SAC} (please refer to Section \ref{sec:measures} for details) for the problem. \vspace{0.5mm}
\item We comprehensively evaluate our method on seven real-world diverse phishing email datasets. The rigorous and extensive experiments show the superiority of our method over the state-of-the-art baselines.
\end{itemize}
\vspace{-0mm}
\section{Related Work}\label{sec:related_work}

\vspace{1mm}
\subsection{\textbf{Phishing attack detection}}
Machine learning has emerged as a natural choice for addressing classification challenges, including the detection of phishing attempts. Over the past decades, extensive research has been dedicated to this area. Numerous machine learning methods have been employed to tackle the issue, creating models capable of classifying content as either benign or phishing \cite{tyagi2018novel, rao2019detection, li2019stacking, zamir2020phishing}. These models leverage classification principles and carefully choose features that represent the content's distinctive characteristics. However, machine learning-based methods are heuristics and require the knowledge of domain experts to manually engineer features from the data that can be outdated and biased. To overcome these limitations, deep learning-based solutions have been proposed for the phishing attack detection problem \cite{xiao2020cnn, abdelnabi2020visualphishnet, yang2019phishing, lin2021phishpedia} and have shown significant advantages over machine learning approaches that use hand-crafted features. AI-based phishing detection methods can predict the vulnerability label (i.e., phishing or benign) of the data (e.g., emails). However, they lack the ability to provide comprehensive explanations for their predictions.

\vspace{-3mm}
\subsection{\textbf{Phishing attack localization}}
Automated deep learning-based techniques for the phishing attack localization problem (i.e., to tackle and improve the explainability
(transparency) of phishing detection) have not yet been well studied. \textit{The interpretable machine-learning research appears to be an appropriate direction for dealing with the phishing attack localization problem}. In short, interpreting approaches (e.g., \cite{caruana2015intelligible, rich2016machine, ribeiro2016should, Chen2018-l2x, bang2021explaining, yoon2019invase, nguyen2021-icvh, jethani2021have}) are used for explaining the behavior of deep learning-based systems or the ground truth label of the data by automatically learning and figuring out the most important information (e.g., attributions or features) existing from the data that are responsible in causing the corresponding decision of black-box models and the ground truth label. \textit{For example, in sentiment analysis, interpretable machine-learning approaches (e.g., \cite{Chen2018-l2x, Vo2023-aim}) help to give a comprehensive explanation for the review (positive or negative) of a movie by figuring out and highlighting the top important keywords or sentences}. In our paper, via addressing phishing attack localization, we not only aim to predict the vulnerability (i.e., phishing or benign) of the emails but also aim to provide a comprehensive interpretation by figuring out and highlighting the most crucial phishing-relevant information causing the emails to be classified as phishing.

\revise{In practice, explaining models can be divided into two categories including "post-hoc explainability techniques" and "intrinsic explainability techniques". Post-hoc explainability techniques (e.g., LIME \cite{RibeiroSG16lime} and SHAP \cite{Lundbergshap}) aim to elucidate the decisions of a black-box model (e.g., a deep learning model, where the internal workings are not easily understandable or interpretable) without modifying the model itself. The techniques are often applied externally to the black-box model to generate explanations specific to its predictions but do not offer a comprehensive understanding of the black-box model's internal architecture or workings. In contrast, intrinsic explainability techniques, a.k.a. self-explanatory models, (e.g., deep neural network-based methods with interpretable components such as L2X \cite{Chen2018-l2x} and AIM \cite{Vo2023-aim}) are integrated directly into a model architecture, providing interpretability as inherent features, and offering explainability as part of their design.}

\revise{It is evident that intrinsic interpretable machine learning methods, a.k.a. self-explanatory models, are strongly suitable for phishing attack localization because they are not only able to make predictions themselves but also automatically learn and figure out the most important information of the data obtained from the models to explain the model's prediction decision. In our paper, we compared the performance of our proposed method with several recent, popular, and state-of-the-art interpretable machine learning approaches falling into the category of intrinsic interpretable models including L2X \cite{Chen2018-l2x}, INVASE \cite{INVASE2019}, ICVH \cite{nguyen2021-icvh}, VIBI \cite{Bang2021-vibi}, and AIM \cite{Vo2023-aim} (please refer to Section \ref{sec:bms} in the experiments section for our brief description of these methods).}

\vspace{0mm}
\section{The proposed approach}\label{sec:framework}

\subsection{The problem statement}\label{sec:problemstatement}
We denote an email data sample as $X=[\mathbf{x}_{1},...\mathbf{x}_{L}]$ consisting of $L$ sentences.
In the scope of our paper, we consider each email as a sequence of sentences. We assume that $X'$s vulnerability label $Y\in\{0,1\}$ (i.e., $1$: phishing and $0$: benign). In the context of phishing attack localization, we aim to build an automatic AI-based approach mainly not only for detecting the vulnerability $Y$ of the email $X$ but also for automatically learning and figuring out the important and phishing-relevant information (i.e., sentences) denoted by $\tilde{X}$ (a subset of $X$) causing $X$ phishing.

It is worth noting that for almost all publicly available phishing-relevant data (e.g., emails), there are only labels related to the data's vulnerability (phishing or benign) by domain experts with the assistance of machine learning or deep learning tools. We almost do not have the ground truth of phishing information (i.e., the information truly causes the data to be classified as phishing). \textit{In the phishing attack localization problem, we name this context as a weakly supervised setting where during the training process}, we only use the vulnerability label $Y$ of the data while not requiring the ground truth of phishing information existing in the data when solving the phishing attack localization problem.

\subsection{Methodology}

\begin{figure}[th]
\vspace{-3mm}
\begin{centering}
\includegraphics[scale=0.66]{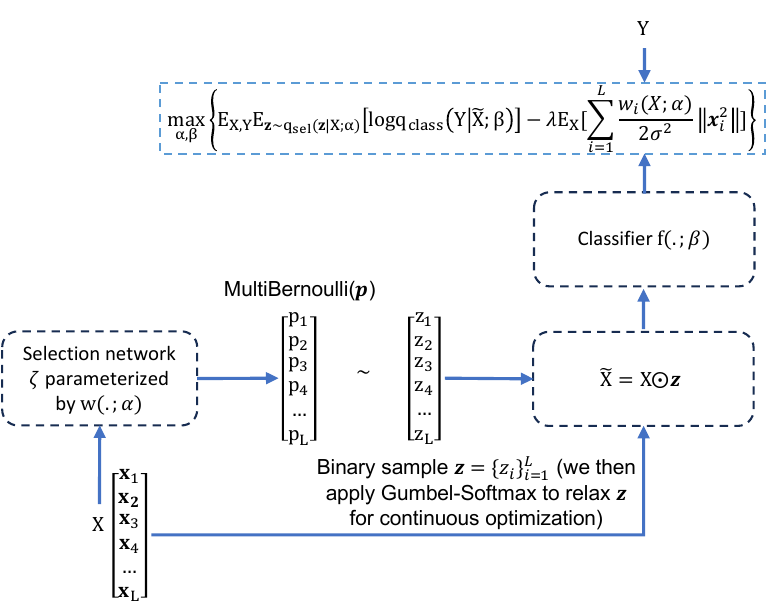}
\par\end{centering}
\vspace{-3mm}
\caption{A visualization of our proposed \ourapp~ method for solving the phishing attack localization problem. \label{fig:proposed_framework}}
\vspace{-3mm}
\end{figure}

We name our proposed method as \ourapp~. Here we present details of how our \ourapp~ method works and addresses the phishing attack localization problem to tackle and improve the explainability
(transparency) of phishing detection. An overall visualization of our method is depicted in Figure \ref{fig:proposed_framework}.

\subsubsection{\textbf{Learning to select the important and phishing-relevant information and the training principle}}

\vspace{-1mm}
\paragraph{\textbf{Phishing-relevant information selection process}}
\vspace{-1mm}

As shown in Figure \ref{fig:proposed_framework}, the first part of our method is the selection network $\zeta$. \textit{It aims to learn and figure out the most important and label-relevant information (i.e., sentences) in each email in an automatic and trainable manner}. Note that, in terms of the phishing email, the key selected information stands for the phishing-relevant information causing the email phishing.

Giving an email $X$ consisting of $L$ sentences from $\mathbf{x}_{1}$ to $\mathbf{x}_{L}$ (i.e., \textit{in practice, each sentence $\mathbf{x}_{i}$ is represented as a vector using a learnable embedding method, please refer to the data processing and embedding section \ref{sec:datape} for details}), to figure out the important and label-relevant sentences $\tilde{X}$ in $X$, we introduce a selection process $\zeta$ (i.e., \textit{it is learnable and maps $\mathbf{R}^L \mapsto [0,1]^L$}) aiming to learn a set of independent Bernoulli latent variables $\mathbf{z}\in\{0,1\}^{L}$ representing the importance of the sentences to the email's vulnerability $Y$. Specifically, each element $z_{i}$ in $\mathbf{z}=\{z_{i}\}_{i=1}^{L}$ indicates whether $\mathbf{x}_{i}$ is related to the vulnerability $Y$ of $X$ (i.e., if  $z_{i}$ is equal to $1$, the sentence $\mathbf{x}_{i}$ plays an important role causing the vulnerability $Y$).

To construct $\mathbf{z}=\{z_{i}\}_{i=1}^{L}$, we model $\mathbf{z} \sim \mathrm{MultiBernoulli}(\mathbf{p}) = \prod_{i=1}^{L}\text{Bernoulli}(p_{i})$, which indicates $\boldsymbol{x}_{i}$ is related to the vulnerability $Y$ of $X$ with probability $p_{i}$ where $p_{i}=\omega_{i}\left(X;\alpha\right)$ with $\omega$ is a neural network parameterized by parameter $\alpha$. It is worth noting that the neural network $\omega$ takes $X$ as input and outputs corresponding $\mathbf{p}=\{p_{i}\}_{i=1}^{L}$. 
With $\mathbf{z}$, we construct $\tilde{X}=\zeta\left(X\right)$ (i.e., the subset sentences that lead to the vulnerability $Y$ of the function $X$) by $\tilde{X}=X\odot \mathbf{z}$, where $\odot$ represents the element-wise product. Note that we can view the selector model as a distribution $q_{sel}(\mathbf{z}|X;\alpha)$ over a selector variable $\mathbf{z}$, which indicates the important features for a given sample of $X$.

\vspace{-0mm}
\paragraph{\textbf{Reparameterization for continuous optimization}}

Recall that the selected sentences $\tilde{X}$ are constructed by $\tilde{X}=\mathbf{z}\odot X$ where $\mathbf{z} \sim \mathrm{MultiBernoulli}(\mathbf{p})$ with $\mathbf{p}$ is the output of the selection process $\zeta$ parameterised by a neural network $\omega(.,\alpha)$. 

To make this computational process (i.e., the process consists of sampling operations from a Multi-Bernoulli distribution) continuous and differentiable during training, we apply the temperature-dependent Gumbel-Softmax trick {\cite{jang2016categorical, MaddisonMT16}} for relaxing each Bernoulli variable $z_{i}$. We sample $z_{i}\left(X;\alpha\right)\sim\text{Concrete}(\omega_{i}(X;\alpha),1-\omega_{i}(X;\alpha))$:

\vspace{-0mm}
\[z_{i}\left(X;\alpha\right)=\frac{\exp\{(\log\omega_{i}+a_{i})/\tau\}}{\exp\{(\log\omega_{i}+a_{i})/\tau\}+\exp\{(\log\left(1-\omega_{i}\right)+b_{i})/\tau\}}
\]
\vspace{-0mm}

where we denote $\omega_{i}\left(X;\alpha\right)$ as $\omega_{i}$ while $\tau$ is a temperature parameter (i.e., that allows us to control how closely a continuous representation from a Gumbel-Softmax distribution approximates this from the corresponding discrete representation from a discrete distribution (e.g., the Bernoulli distribution)), random noises $a_{i}$ and $b_{i}$ independently drawn from \textbf{Gumbel} distribution
$G = - \log(- \log u)$ with $\ u \sim \textbf{Uniform}(0,1)$.

\vspace{-0mm}
\paragraph{\textbf{Mutual information for guiding the selection process}}

In information theory \cite{TheoryofC, EofIT2006}, mutual information is used to measure the mutual dependence between two random variables. In particular, it quantifies the information obtained about one random variable by observing the other. To illustrate, consider a scenario where A denotes the outcome of rolling a standard 6-sided die, and B represents whether the roll results in an even number (0 for
even, 1 for odd). Evidently, the information conveyed by B provides insights into the value of A, and vice versa. In other words, these random variables possess mutual information. 

Leveraging this property of mutual information and inspired by \cite{learning-to-explain-l2x,nguyen2021-icvh}, we maximize the mutual information between $\tilde{X}$ and $Y$ as mentioned in Eq. (\ref{eq:max_info}) with the intuition is that by using the information from $Y$, the selection process $\zeta$ will be learned and enforced to obtain the most meaningful $\tilde{X}$ (i.e., $\tilde{X}$ can predict the vulnerability $Y$ of $X$ correctly).

By viewing $\tilde{X}$ and $Y$ as random variables, the selection process (model) $\zeta$ is learned by maximizing the mutual information between $\tilde{X}$ and $Y$ as follows:

\vspace{-0mm}
{
\begin{equation}
\max_{\zeta}\,\mathbf{I}(\tilde{X},Y)\label{eq:max_info}
\end{equation}
}
\vspace{-0mm}

Following \cite{EofIT2006}, we expand Eq. (\ref{eq:max_info}) further as the Kullback-Leibler divergence (i.e., it measures the relative entropy or difference in information represented by two distributions) of the product of marginal distributions of $\tilde{X}$ and $Y$ from their joint distribution:

\vspace{-0mm}
{
\begin{align}
\mathbf{I}(\tilde{X},Y)= & \int p(\tilde{X},Y)\log\frac{p(\tilde{X},Y)}{p(\tilde{X})p(Y)}d\tilde{X}dY\nonumber \\ 
\geq & \int p(Y,\tilde{X})\log\frac{q(Y\mid\tilde{X})}{p(Y)}dYd\tilde{X}\label{eq:1sm}
\end{align}
}
\vspace{-0mm}

In practice, estimating mutual information is challenging as we typically only have access to samples but not the underlying distributions. Therefore, in the above derivation, we apply to use a variational distribution $q(Y|\tilde{X})$ to approximate the posterior $p(Y\mid\tilde{X})$, hence deriving a variational lower bound of $\mathbf{I}(\tilde{X}, Y)$ for which the equality holds if $q(Y\mid\tilde{X})=p(Y\mid\tilde{X})$. This can be further expanded as:

\vspace{-0mm}
{
\begin{align}
\mathbf{I}(\tilde{X},Y)\geq & \int p(Y,\tilde{X},X)\log\frac{q(Y\mid\tilde{X})}{p(Y)}dYd\tilde{X}dX\nonumber \\
= & \mathbf{E}_{X,Y}\mathbf{E}_{\tilde{X}|X}[\log q(Y|\tilde{X})]+\text{const}\label{eq:2sm}
\end{align}
}

To model the conditional variational distribution $q(Y|\tilde{X})$, we introduce a classifier implemented with a neural network $f(\tilde{X};\beta)$, which takes $\tilde{X}$ as input and outputs its corresponding label $Y$ (i.e., \textit{we view the classifier model as a distribution $q_{class}(Y|\tilde{X};\beta)$}). Our objective is to learn the selection process (model) as well as the classifier to maximize the following objective function:

{
\begin{equation}
\underset{\alpha,\beta}{\text{max}}\left\{\mathbf{E}_{X,Y}\mathbf{E}_{\mathbf{z}\sim q_{sel}(\mathbf{z}|X;\alpha)}[\text{log}q_{class}(Y|X\odot\mathbf{z};\beta)]\right\}\label{eq:max_info_mutual_v2}
\end{equation}
}

The mutual information facilitates a joint training process for the classifier and the selection process. \textit{The classifier learns to identify a subset of features leading to a data sample's label while the selection process is designed to select the best subset of features according to the feedback of the classifier.}

\subsubsection{\textbf{Benefits as well as potential weaknesses of the mutual information training principle and our innovative solutions}}\label{sec:bandw}

With the training principle mentioned in Eq. (\ref{eq:max_info}), we aim to maximize the mutual information between $\tilde{X}$ and $Y$ for guiding the whole training process to figure out the sentences related to an email's vulnerability. Recall that during this training process, the classifier learns to identify a subset of sentences leading to an email sample's vulnerability label while the selection process is designed to select the best subset according to the feedback of the classifier.

\textit{\textit{In short, this joint training process between the classifier $f(.,\beta)$ and the selection network $\omega(.,\alpha)$ brings benefits for selecting the important and phishing-relevant sentences from phishing email; However, we observe two potential limitations of this training principle as follows:}}

\paragraph{\textbf{Obtaining a superset of phishing-relevant sentences}}
\textit{The first limitation of the training principle mentioned in Eq. (\ref{eq:max_info}) is that it does not theoretically guarantee to eliminate sentences unrelated to the vulnerability of a specific email}. Therefore, the set of selected sentences can be a superset of the true phishing-relevant sentences. In the worst case, a selection process can always select all the sentences in an email, which is still a valid solution for the above maximization mentioned in Eq. (\ref{eq:max_info}). To deal with this problem to let the model be able to successfully select and highlight true phishing-relevant information, inspired by \cite{tishby2000information,slonim2000agglomerative} about using information bottleneck theory, we propose an additional term for training the selection process (model) $\zeta$, derived from the following principle:

{\small{}
\begin{equation}
\max_{\zeta}\,(\mathbf{I}(\tilde{X},Y)-\lambda\mathbf{I}(X,\tilde{X}))\label{eq:max_info-2}
\end{equation}
}{\small\par}

\noindent where $\lambda$ is a hyper-parameter indicating the weight of the second mutual information.

\textit{By minimizing the mutual information between $X$ and $\tilde{{X}}$, we encourages $\tilde{{X}}$ to be as "different" to $X$ as possible}. In other words, the selection process prefers to select a smaller subset that excludes the sentences unrelated to vulnerability $Y$ of the corresponding data $X$. Accordingly, we can derive an upper bound of the minimization between mutual information between $X$ and $\tilde{X}$:\vspace{-2mm}

{\small{}
\begin{gather}
\mathbf{I}(\tilde{X},X)=\int p(\tilde{X},X)\log\frac{p(\tilde{X}|X)}{p(\tilde{X})}d\tilde{X}dX\nonumber \\
\leq\mathbf{E}_{X}\mathbf{E}_{\tilde{X}|X}[\log\frac{p(\tilde{X}|X)}{r(\tilde{X})}]\label{eq:3sm}
\end{gather}
}{\small\par}

\vspace{-0mm}

for any distribution $r(\tilde{X})$.

\vspace{1mm}
We then further derive $\mathbf{I}(\tilde{X}, X)$ as the Kullback-Leibler divergence of the product of marginal distributions of $\tilde{X}$ and $X$ from their joint distribution:

{
\begin{gather}
\mathbf{E}_{X}\mathbf{E}_{\tilde{X}|X}[\log\frac{p(\tilde{X}|X)}{r(\tilde{X})}]=\mathbf{E}_{X}\mathbf{E}_{\tilde{X}|X}[\sum_{i=1}^{L}\log\frac{p(\tilde{\boldsymbol{x}_{i}}|X)}{r(\tilde{\boldsymbol{x}_{i}})}]\nonumber \\
\,\,\,\,\,\,\,\,\,\,\,\,\,\,=\sum_{i=1}^{L}\mathbf{E}_{X}[D_{KL}(p(\tilde{\boldsymbol{x}_{i}}|X)\Vert r(\tilde{\boldsymbol{x}_{i}}))]\label{eq:4sm}
\end{gather}
}\vspace{1mm}

Minimizing $\mathbf{I}(\tilde{X},X)$ is now equivalent to minimizing the KL divergence between $p(\tilde{\boldsymbol{x}_{i}}|X)$ and $r(\tilde{\boldsymbol{x}_{i}})$.
Therefore, one can view $r(\tilde{\boldsymbol{x}_{i}})$ as the prior distribution, which is constructed by $r(\tilde{\boldsymbol{x}_{i}})=\mathcal{N}(\tilde{\boldsymbol{x}}_{i}|0,\sigma^{2})$. Given the fact that $p(\tilde{\boldsymbol{x}}_{i}|X)$ is a Gaussian mixture distribution (i.e., between $p_{i}\mathcal{N}(\tilde{\boldsymbol{x}}_{i}\mid\boldsymbol{x}_{i},\sigma^{2})$ and ($1-p_{i})\mathcal{N}(\tilde{\boldsymbol{x}_{i}}\mid0,\sigma^{2})$ where $\sigma>0$ is a small number), the intuition is that the prior prefers the small values centered at $0$. In this way, $p(\tilde{\boldsymbol{x}}_{i}|X)$ is encouraged to select fewer sentences. Moreover, the KL divergence $D_{KL}(p(\tilde{\boldsymbol{x}}_{i}|X)\Vert r(\tilde{\boldsymbol{x}}_{i}))$ can be computed by the following approximation \cite{gal2016dropout}:\vspace{-1mm}

{
\begin{gather}
\frac{\omega_{i}(X;\alpha)}{2\sigma^{2}}\left\Vert \boldsymbol{x}_{i}^{2}\right\Vert +(\log\,\sigma+\frac{1}{2}\sigma^{2})+\text{const}
\end{gather}
}

Combining $\mathbf{I}(\tilde{X},Y)$ and $\mathbf{I}(X,\tilde{X})$ by $\max\,(\mathbf{I}(\tilde{X},Y)-\lambda\mathbf{I}(X,\tilde{X}))$, we get a unified training objective:

{
\begin{gather}
\max_{\alpha,\beta}\{\mathbf{E}_{X,Y}\mathbf{E}_{\mathbf{z}\sim q_{sel}(\mathbf{z}|X;\alpha)}[\text{log}q_{class}(Y|X\odot\mathbf{z};\beta)]\nonumber \\
\,\,\,\,\,\,\,\,\,\,\,\,\,\,\,\,\,\,\,\,\,-\lambda\mathbf{E}_{X}[\sum_{i=1}^{L}\frac{\omega_{i}\left(X;\alpha\right)}{2\sigma^{2}}\left\Vert \boldsymbol{x}_{i}^{2}\right\Vert\}\label{eq:op_2_2}
\end{gather}
}

\vspace{2mm}
\paragraph{\textbf{Encoding the vulnerability label via its selections instead of via truly meaningful information}}
The selected features obtained from the joint training process of the classifier $f(.,\beta)$ and the selection network $\omega(.,\alpha)$ cause the other potential limitation for the training principle mentioned in Eq. (\ref{eq:max_info}). In particular, the predictions of the classifier $f(.,\beta)$ can be based more on the features selected from the selection network $\omega(.,\alpha)$ than the underlying information contained in the features. In this case, the selected information (i.e., sentences) can be any subsets of the entire sentences and can be less likely to be meaningful ones from the data.

To deal with this problem and ensure the learnable selection process respecting the data distribution to select the meaningful and label-relevant information (e.g., sentences) of the data, in addition to learning the classifier jointly with the selection network as mentioned in Eq. (\ref{eq:op_2_2}), inspired by \cite{realx2021}, we propose to learn the classifier model $f(.,\beta)$ disjointly to approximate the ground truth conditional distribution of $Y$ given  
$X_R$ where $X_{R}=X\odot \mathbf{r}$ with $\mathbf{r}\sim \text{MultiBernoulli}(0.5)$ (denoted by $\mathbf{r}\sim \text{B}(0.5)$ for short). \textit{This procedure helps adjust the classifier to let it not only be affected by the information obtained from the selection network but also based on the information from the data to update its parameters}. That helps prevent the problem of encoding the vulnerability label via its selections to improve the data representation learning process. We name this procedure as a \textbf{data-distribution mechanism}. This procedure is then expressed as learning $q_{class}(.;\beta)$ to maximize:

\vspace{-0mm}
{
\begin{gather}
\mathbf{E}_{X,Y}\mathbf{E}_{\mathbf{r}\sim \text{B}(0.5)}\{\text{log}q_{class}(Y|X\odot\mathbf{r};\beta)\}\label{eq:realx}
\end{gather}
}
\vspace{-0mm}

To make this computational procedure (i.e., it consists of sampling operations from a Multi-Bernoulli distribution) continuous and differentiable during the training process, we apply to use the temperature-dependent Gumbel-Softmax trick {\cite{jang2016categorical, MaddisonMT16}} for relaxing each Bernoulli variable $r_{i} \in \mathbf{r}$ using the RelaxedBernoulli distribution function from \cite{RelaxedBernoulli}. \textit{Noting that in Eq.(\ref{eq:realx}), by setting $\mathbf{r}\sim \text{B}(0.5)$ to randomly select sentences from each email instead of using the entire email content, we also aim to introduce randomness into the training corpus, akin to data augmentation, which helps enhance the generalization capability of the selection network that facilitates the performance of the classifier.}

\vspace{0mm}
\subsubsection{\textbf{A summary of our \ourapp~ method}}
Algorithm 1 shows the details of our proposed \ourapp~ method in the training phase. \textit{It is worth noting that during the training process, our model is trained to learn and figure out the most important and phishing-relevant sentences in corresponding emails without using any information about the ground truth of phishing-relevant sentences}. This shows a great advantage of using our method for phishing attack localization in real-world scenarios because in practice, to almost all publicly available email datasets, there is only information about the vulnerability $Y$ (i.e., phishing or benign) of the data by domain experts with the help of machine learning and deep learning tools.

\vspace{-1mm}
\paragraph{\textbf{The inference (testing) phase}}

After the training phase, the selection network $\zeta$ is capable of selecting the most important and phishing-relevant sentences of a given email data $X$ by offering a high value for the corresponding coordinates
$\omega_{i}\left(X;\alpha\right)$, meaning that $\omega_{i}\left(X;\alpha\right)$ represents the influence level of the sentence $\boldsymbol{x}_{i}$. We hence can pick out the most relevant sentences based on the magnitude of $\omega_{i}\left(X;\alpha\right)$. Using the selected information, the trained classifier then can predict the vulnerability of the associated email data. \textit{In our paper, with the aim of providing the most highly qualified and concise explanation of the vulnerability of email data to users,} \textbf{\textit{we primarily assess the model's performance based on the most important (top-1) selected sentence from each email.}}

\vspace{0mm}
\RestyleAlgo{ruled}
\begin{algorithm*}[ht]
\DontPrintSemicolon 
\LinesNumbered

\vspace{1mm}
\KwIn{
An email dataset $S=\left\{ \left(X_{1},Y_{1}\right),\dots,\left(X_{N_{S}},Y_{N_{S}}\right)\right\} $ where each email $X_{i}=[\mathbf{x}_{1},...\mathbf{x}_{L}]$ consisting of $L$ sentences while its vulnerability label $Y_{i}\in\left\{ 0,1\right\} $ (i.e., $1$: phishing and
$0$: benign). \newline
We denote the number of training iterations $nt$; the mini-batch size $m$;
the trade-off hyper-parameter $\lambda$. \newline
We randomly partition $S$ into three different sets including the training set $S_{train}$ (for training the model), the validation set $D_{val}$ (for saving the model during the training process), and the testing set $D_{test}$ (for evaluating the model performance).
}
\BlankLine
We initialize the parameters $\alpha$ and $\beta$ of the selection model $\zeta$ (i.e., parameterized by $\omega(.,\alpha)$) and the classifier model $f(.,\beta)$, respectively.
\BlankLine
\For{$t=1$ to $nt$}
{
Choose a mini-batch of embedded phishing-relevant emails denoted by $\{(X_{i},Y_{i})\}_{i=1}^{m}$.
\BlankLine
Update the classifier's parameter $\beta$ via minimizing the following cross-entropy loss $\mathcal{L}_{ce} \mathbb{E}_{X,Y}\mathbb{E}_{\mathbf{r}\sim B(0.5)}[\mathcal{L}_{ce}(Y,f_{\beta}(X\odot\mathbf{r})]$ using the Adam optimizer \cite{KingmaB14}. Note that minimizing this function is equivalent to maximizing the objective function in Eq. (\ref{eq:realx}).\;
\BlankLine
Update the classifier's parameter $\beta$ and the selection model parameter's $\alpha$ via minimizing the following objective function $\mathbb{E}_{X,Y}\mathbb{E}_{\mathbf{z}\sim q_{sel}(\mathbf{z}|X;\alpha)}[\mathcal{L}_{ce}(Y,f_{\theta}(X\odot\mathbf{z}))]+\lambda\mathbb{E}_{X}[\sum_{i=1}^{L}\frac{\omega_{i}\left(X;\alpha\right)}{2\sigma^{2}}\left\Vert \boldsymbol{x}_{i}^{2}\right\Vert]$ using the Adam optimizer.\;
\BlankLine
}
\BlankLine
\vspace{1mm}
\KwOut{The trained model for phishing attack localization.}
\vspace{1mm}
\caption{The algorithm of our proposed \ourapp~ method for the phishing attack localization problem.\label{alg:The-training-algorithm-prom}}
\end{algorithm*}
\vspace{0mm}
\vspace{2mm}
\section{Experiments}\label{sec:experiments}

\subsection{Experimental Designs}
\revise{The fundamental goal of this experiment section is to evaluate our proposed \ourapp~ method and compare it with several state-of-the-art baseline approaches for phishing attack localization. Below, we present the main research questions of our paper.}

\vspace{2mm}
\textbf{(RQ1) \rqone}

Research on phishing detection is one of the most popular solutions for defeating phishing attacks. There have been many machine learning-based and deep learning-based approaches proposed to predict the vulnerability (i.e., phishing or benign) of the associated data (e.g., emails). \textit{Although achieving promising performances, AI-based phishing detection methods lack an explanation of the vulnerability prediction (i.e., especially in identifying the specific information causing the data to be classified as phishing)}. To this end, in this paper, we propose an innovative deep learning-based method for phishing attack localization on email data (the most common phishing way) where we not only aim to detect the vulnerability of the data but also learn and figure out the important and phishing-relevant information existing in the phishing data to provide corresponding comprehensive explanations (i.e., about the key information causing the data phishing). That helps tackle the lack of explainability (transparency) existing in phishing email attack detection. We compare the performance of our \ourapp~ method with several state-of-the-art baselines from explainable machine learning.

Via this research question, we mainly aim to investigate the effectiveness of our \ourapp~ method and baselines in solving the phishing attack localization problem. \textit{In particular, we evaluate their capability in predicting each email's vulnerability and identifying the most important (top-1) phishing-relevant sentence for offering the most concise and highly qualified explanation.} The selected sentence in each email can also reflect the human cognitive principles/triggers (i.e., the definition of "principles of influence", namely Reciprocity, Consistency, Social Proof, Authority, Liking, and Scarcity) often used in phishing emails (e.g., \cite{Akbar2014, Butavicius2015, Ferreira2015, Cognitive2019}).

\vspace{2mm}
\textbf{(RQ2) \rqtwo}

As studied and pointed out in phishing attack-relevant research \cite{Akbar2014, Butavicius2015, Ferreira2015, Cognitive2019}, among cognitive triggers used in phishing attacks, some of them (Authority, Scarcity, and Consistency) are more popular than others (Reciprocity, Social, Proof, and Liking). Therefore, in this research question, we investigate how our \ourapp~ method compared to the baselines in consisting of more important and popular cognitive triggers used in phishing emails in the most important (top-$1$) selected sentences.

\vspace{3mm}
\textbf{(RQ3) \rqthree}

\vspace{1mm}
The training principle in Eq. (\ref{eq:max_info}) induced from the mutual information between the selected information and the corresponding label brings benefits for selecting the important and phishing-relevant sentences from phishing emails, however, as mentioned in Section \ref{sec:bandw}, it has two potential limitations regarding \textit{"obtaining a superset of phishing-relevant sentences"} and \textit{"encoding the vulnerability label via its selections instead of via truly meaningful information"}. 

To this end, we further propose two innovative solutions including the information bottleneck theory training term and data-distribution mechanism. In short, our innovative information bottle-neck training mechanism ensures that only important and label-relevant information (i.e., sentences) will be kept and selected while the data-distribution mechanism ensures the learnable selection process respecting the data distribution to select the meaningful and label-relevant information. It also aids in improving the generalization capability of the selection network, thereby enhancing the performance of the classifier through a random selection process that incorporates data augmentation (please refer to Section \ref{sec:bandw} for our detailed explanation).

In this research question, we aim to assess the effectiveness of our proposed information bottleneck theory training term (mentioned in Eq.(\ref{eq:op_2_2})) and data-distribution mechanism (mentioned in Eq. (\ref{eq:realx})) in improving the capability of our \ourapp~ method in the process of learning and figuring out the most important and truly phishing-relevant information (i.e., sentences) in phishing-relevant emails.

\vspace{-0mm}
\subsubsection{\textbf{Studied datasets}} \label{sec:studied_datasets}

We conducted the experiments on seven diverse phishing-relevant email datasets including \textbf{IWSPA-AP} (i.e., the dataset was collected as part of a shared task to try and address phishing scam emails), \textbf{Nazario Phishing Corpus} (i.e., the dataset of phishing emails (received by one user) that are surrounded by HTML code and need to be stripped to get the useful text data), \textbf{Miller Smiles Phishing Email} (i.e., the dataset contains the bodies of the phishing email scams), \textbf{Phish Bowl Cornell University} (i.e., the dataset contains phishing emails that have been spotted and reported by students and staff at Cornell University), \textbf{Fraud emails} (i.e., the dataset contains fraudulent emails attempting Nigerian Letter where all the emails are in one text file and contain a large amount of header data), \textbf{Cambridge} (i.e., the data set contains a large number of email headers (involving information such as sending and receiving addresses and email subjects) and the body content of phishing emails), and \textbf{Enron Emails} (i.e., the email dataset was released as part of an investigation into Enron, which consists of emails from mostly senior management of Enron).

These datasets contain the common email header and email body. We found that most of the email data are stored in the form of webpages on public websites. We re-processed the datasets before using them for the training and testing processes in terms of removing non-ASCII characters and unrelated information (e.g., HTML language syntax). \revise{We also removed duplicated emails in the used datasets as well as similar emails (when they are in the same label category, i.e., phishing or benign).
In the end, we utilized around 40,000  email data samples, where half of them are phishing, from the datasets to conduct our comprehensive and rigorous experiments. Note that the email data samples used in our experiments cover a wide spectrum of writing styles, tones, and structures commonly encountered in email communications (e.g., formal business correspondences, informal personal messages, transactional emails (such as order confirmations or account notifications), marketing emails promoting products or services, and automated responses to specific actions or events).}

\subsubsection{\textbf{Data processing and embedding}}\label{sec:datape}

We preprocessed the datasets before injecting them into deep neural networks of our \ourapp~ method and baselines. In the context of our paper, we view each email as a sequence of sentences and aim to learn and figure out the most important sentence in each email contributing to the email's vulnerability label (phishing or benign).

We embedded sentences of each email into vectors. For instance, consider the following sentence \emph{"Your account has been suspended."}, to embed this sentence, we tokenized it to a sequence of tokens including \emph{"your"}, \emph{"account"}, \emph{"has"}, \emph{"been"}, \emph{"suspended"}, and \emph{"."} using the common Natural Language Toolkit (NLTK). We then used a 150-dimensional token Embedding layer followed by a Dropout layer with a dropped fixed probability $p=0.2$, a 1D convolutional layer with the filter size $150$ and kernel size $3$, and a 1D max pooling layer to encode each sentence. Finally, a mini-batch of emails in which each email consisting of $L$ encoded sentences was fed to our proposed \ourapp~ method and the baselines.

In our paper, the length of each email is padded or truncated with $L=100$ sentences (i.e., we base on the quantile values of the emails\textquoteright{} length of the used datasets to decide the length of each email). We observe that almost all important information relevant to the phishing vulnerability lies in the $100$ first sentences or even lies in some very first sentences.

\vspace{-0mm}
\subsubsection{\textbf{Measures}}\label{sec:measures}

To the best of our knowledge, our method is one of the very first approaches proposed for solving the phishing attack localization problem aiming to tackle and improve the explainability (transparency) of phishing detection. \textit{Therefore, studying appropriate measures for phishing attack localization is necessary and also is one of the contributions of our work.} 

To evaluate the performance of our \ourapp~ method and the baselines in phishing attack localization, based on the explainable machine learning and email phishing domain knowledge, we introduce two main metrics to measure the quality of the most important selected sentence in each phishing-relevant email including \textbf{Label-Accuracy} and \textbf{Cognitive-True-Positive}. 

Via the \textbf{Label-Accuracy} metric, we measure whether the selected sentences obtained from each model can accurately predict the true vulnerability label (i.e., phishing or benign) of the associated email. \textit{The intuition is that the most important sentences contribute the most to the email's vulnerability, especially to phishing emails.} In our experiments, we assess each model's top-1 selected sentence and measure if this sentence can effectively predict the email's vulnerability without considering all sentences in the email. The higher the value of the Label-Accuracy measure, the better the model's performance in figuring out and selecting crucial and label-relevant information from the data.

For the \textbf{Cognitive-True-Positive} metric, we aim to investigate if the top-$1$ selected sentence from each method also reflects the human cognitive principles (triggers) used in phishing emails (i.e., Reciprocity, Consistency, Social Proof, Authority, Liking, and Scarcity \cite{Akbar2014, Butavicius2015, Ferreira2015, Cognitive2019} based on the associated keywords). \textit{In particular, in our experiments, we consider the most important (top-$1$) selected sentences of phishing emails and calculate how many percent of these selected sentences consist of cognitive triggers.}

It is worth noting that to measure if the top-$1$ selected sentence obtained from each method reflects and captures the human cognitive principles (triggers) used in each phishing email, we are particularly based on the keywords and phrases often used in each cognitive principle. To obtain the keywords and phrases, for each cognitive principle, we first base on its definition (descriptions) pioneered in the well-known work \cite{Cialdini1984} (widely cited and used in phishing-related studies \cite{Akbar2014, Butavicius2015, Ferreira2015, Cognitive2019}). We then use ChatGPT \cite{ChatGPT} to obtain all the possible keywords and phrases related to each cognitive principle’s definition (descriptions). For example, to the Scarcity principle, some of the related keywords and phrases can be relevant to Time-sensitive language (e.g., "Act now," or "Expires soon"), Limited availability (e.g., "Only a few left"), and Threats of consequences (e.g., "Will be deleted" or "Will lose access"). By relying on the definition (descriptions) of each cognitive principle with the help of ChatGPT in finding all the possible and main keywords and phrases, related to the definition (descriptions) of each cognitive principle, used in the Cognitive-True-Positive measure, we ensure this process is reliable and objective.

Among cognitive triggers used in phishing attacks, some of them (Authority, Scarcity, and Consistency) are more popular than others (Reciprocity, Social, Proof, and Liking). To investigate how our \ourapp~ method compared to the baselines in consisting of more important and popular cognitive triggers via the most important selected information. We introduce the \textbf{SAC} measure, standing for the measure related to Scarcity, Authority, and Consistency of cognitive principles. \textit{In particular, we consider the top-1 selected sentences of phishing emails and calculate how many percent of these selected sentences consist of the commonly used cognitive triggers.}

\vspace{0mm}
\subsubsection{\textbf{Baseline methods}}\label{sec:bms}

\revise{The baselines of our method are recent, popular, and state-of-the-art interpretable machine learning approaches falling into the category of intrinsic interpretable models including L2X \cite{Chen2018-l2x}, INVASE \cite{INVASE2019}, ICVH \cite{nguyen2021-icvh}, VIBI \cite{Bang2021-vibi}, and AIM \cite{Vo2023-aim} that we apply to solve the phishing attack localization problem (i.e., where we not only aim to detect the vulnerability label (phishing or benign) of phishing-related data but also learn and figure out the most important and phishing information existing in the data to provide useful and concise explanations about the corresponding phishing vulnerability).} \revise{Intrinsic interpretable machine learning techniques, a.k.a. self-explanatory models, are strongly suitable for phishing attack localization because they are able to not only make predictions themselves but also figure out the most important features of the data obtained from the model to explain the model's predictive decision (i.e., please refer to section (\ref{sec:related_work}) for our brief description about intrinsic interpretable techniques).}

\vspace{1mm}
We briefly summarize the baselines as follows:
\begin{itemize}
\vspace{0mm}
    \item \textbf{L2X} \cite{Chen2018-l2x}. An efficient instance-wise feature selection method leverages mutual information for model interpretation. L2X aims to extract a subset of features that are most informative for each given example and the associated model prediction response.
    \vspace{0.5mm}
    \item \textbf{INVASE} \cite{INVASE2019}. Another effective approach for instance-wise feature selection. INVASE consists of three neural networks including a selector network, a predictor network, and a baseline network which are used to train the selector network using the actor-critic methodology.
    \vspace{0.5mm}
    \item \textbf{ICVH} \cite{nguyen2021-icvh}. One of the first interpretable deep learning-based methods applied for source code vulnerability localization. ICVH is based on mutual information and a multi-Bernoulli distribution selection process for selecting vulnerability-relevant source code statements.
    \vspace{0.5mm}
    \item \textbf{VIBI} \cite{Bang2021-vibi}. VIBI is a system-agnostic method providing a brief and comprehensive explanation by adopting an information-theoretic principle, the information bottleneck principle, as a criterion for finding such explanations.
    \vspace{0.5mm}
    \item \textbf{AIM} \cite{Vo2023-aim}. A recent innovative additive instance-wise framework for model interpretation. AIM integrates both feature attribution (producing relative importance scores to each feature) and feature selection (directly identifying the subset of features most relevant to the model behavior being explained) into an effective framework for multi-class model interpretation.
\end{itemize}

\vspace{0mm}
It is worth noting that our proposed \ourapp~ method and the used baselines can work on the weakly supervised setting where during the training process, we only utilize the vulnerability label of the data, without requiring the ground truth of phishing information in the data for the task of phishing attack localization.

\vspace{0mm}
\subsubsection{\textbf{Model's configuration}}\label{sec:modelc}

For the \textbf{L2X} \cite{Chen2018-l2x}, \textbf{ICVH} \cite{nguyen2021-icvh}, \textbf{VIBI} \cite{Bang2021-vibi}, and \textbf{AIM} \cite{Vo2023-aim} methods, they were proposed to explain the output of black-box learning models.
To use these methods for phishing attack localization, we keep their principles and train the models to directly approximate $p(Y\mid X)$ using $p(Y\mid\tilde{X})$ where $\tilde{X}$ consists of the selected sentences while $Y$ is ground truth label of the data $X$ instead of the output from the black-box model. 

\textbf{Note that} the L2X, ICVH, VIBI, and AIM methods were also proposed to work with sequential text data. Therefore, to these methods, for the architecture of the selection network obtaining $\tilde{X}$ as well as the classifier working on $\tilde{X}$ to mimic $p(Y\mid X)$, we follow the structures used in the corresponding original papers with the same suggested value ranges for hyper-parameters. For the INVASE method, because it was originally designed for working with tabular data, to let it be able to be applicable for sequential text data, we keep its principle and use the selection network as in our \ourapp~ method.

We implemented our \ourapp~ method in Python using Tensorflow \cite{abadi2016tensorflow}. The trade-off parameter $\lambda$ is in $\{10^{-1},10^{-2},10^{-3}\}$ while $\sigma$ is in $\{10^{-1},2\times10^{-1},3\times10^{-1}\}$. For the networks $\omega\left(\cdot;\alpha\right)$ and $f\left(\cdot;\beta\right)$, we used deep feed-forward neural networks having three and two hidden layers with the size of each hidden layer in $\left\{100,300\right\}$, respectively. The dense hidden layers are followed by a ReLU function as nonlinearity and Dropout \cite{srivastava14a} with a retained fixed probability $p=0.8$ as regularization. The last dense layer of the network $\omega\left(\cdot;\alpha\right)$ for learning a discrete distribution is followed by a sigmoid function while the last dense layer of the network $f\left(\cdot;\beta\right)$ is followed by a softmax function for predicting. The temperature $\tau$ for the Gumbel softmax distribution is in $\{0.5,1.0\}$. Note that we utilized the commonly used values for these hyper-parameters.

For our \ourapp~ method and baselines, we employed the Adam optimizer \cite{KingmaB14} with an initial learning rate of $10^{-3}$, while the mini-batch size is $128$. We split the data set into three random partitions. The first partition contains $80\%$ for training, the second partition contains $10\%$ for validation and the last partition contains $10\%$ for testing. We used $10$ epochs for the training process. We additionally applied gradient clipping regularization to prevent over-fitting. For each method, we ran the corresponding model several times and reported the averaged \textbf{Label-Accuracy}, \textbf{Cognitive-True-Positive}, and \textbf{SAC} measures. We ran our experiments on a 13th Gen Intel(R) Core(TM) i9-13900KF having 24 CPU Cores at 3.00 GHz with 32GB RAM, integrated Gigabyte RTX 4090 Gaming OC 24GB. The source code and data samples for reproducing the experiments of our \ourapp~ method are published at \textit{\url{https://anonymous.4open.science/r/AI2TALE/}}.

\subsection{Experimental results}\label{sec:exp_results}

\textbf{RQ1: \rqone}

\vspace{-1mm}
\paragraph{\textbf{Approach}}

We compare the performance of our \ourapp~ method with the baselines including \textbf{L2X} \cite{Chen2018-l2x}, \textbf{INVASE} \cite{INVASE2019}, \textbf{ICVH} \cite{nguyen2021-icvh}, \textbf{VIBI} \cite{Bang2021-vibi}, and \textbf{AIM} \cite{Vo2023-aim} in the task of phishing attack localization (in terms of \textit{not only predicting the vulnerability of the email data but also learning and figuring out the most important (top-1) and phishing-relevant information existing in the data}. The selected information helps provide useful and concise explanations about the vulnerability of the phishing email data for the users) using the \textbf{Label-Accuracy} and \textbf{Cognitive-True-Positive measures}.

\vspace{-1mm}
\paragraph{\textbf{Quantitative Results}}

The experimental results in Table \ref{tab:my_label1} show that our \ourapp~ method obtains the best performance on both Label-Accuracy and Cognitive-True-Positive compared to the baselines. Importantly, our method achieves a significantly higher performance, with improvements ranging from approximately 1.5\% to 3.5\% compared to state-of-the-art baselines, measured by the combined average performance of two main metrics, i.e., Label-Accuracy and Cognitive-True-Positive. The results demonstrate the effectiveness and advancement of our method for phishing attack localization in learning and figuring out the most meaningful and crucial information leading to the vulnerability of the email data, especially for phishing ones, compared to the baselines.

\begin{table}[th]
    \centering{}
    \vspace{1mm}
    \caption{The performance of our \ourapp~ method and the baselines for the Label-Accuracy (Label-Acc) and Cognitive-True-Positive (Cognitive-TP) measures, as well as their combined average results (denoted as Average), on the testing set (the best results in \textbf{bold}).}
    \label{tab:my_label1}
    \vspace{-3mm}
    \resizebox{0.95\columnwidth}{!}{
    \begin{tabular}{c|c|c|c}
    \hline
    \textbf{Methods} & \textbf{Label-Acc} & \textbf{Cognitive-TP} & \textbf{Average} \\
    \hline
    INVASE \cite{INVASE2019} & 98.30\% & 97.20\% & 97.75\%\\
    \hline
    ICVH \cite{nguyen2021-icvh} &	96.72\%	& 98.10\% & 97.41\% \\
    \hline
    L2X \cite{Chen2018-l2x} &	98.25\%	& 97.20\% & 97.73\% \\
    \hline
    VIBI \cite{Bang2021-vibi} & 96.65\% & 94.99\% & 95.82\% \\
    \hline
    AIM \cite{Vo2023-aim} & 98.40\% & 97.10\% & 97.75\% \\
    \hline
    \multirow{2}{*}{\ourapp~ (Ours)} & \multirow{2}{*}{\textbf{99.33\%}} & \multirow{2}{*}{\textbf{98.95\%}} & {\textbf{99.14\%}} \tabularnewline
    &  &  & \textcolor{blue}{$\uparrow$ $\sim($1.5\% $\rightarrow$3.5\%)} \tabularnewline
    \hline
    \end{tabular}}
    \vspace{-5mm}
\end{table}

In addition, the experimental results on the Cognitive-True-Positive measure of our \ourapp~ method and the baselines shown in Table \ref{tab:my_label1} also indicate that the most important (top-1) selected sentences from our method and the baselines also reflect the cognitive triggers used in phishing emails. Compared to the baselines, our proposed \ourapp~ method achieves the best performance on the Cognitive-True-Positive measure. 

\begin{table}[th]
\vspace{1mm}
\caption{The ground truth and the predicted label based on the most important (top-1) selected sentence (highlighted in yellow) of each email from our \ourapp~ method.}
\label{tab:visualizationresult1}
\vspace{-3mm}
\centering
\resizebox{0.96\linewidth}{!}{%
\begin{tabular}{p{1cm}|p{1cm}|p{8.8cm}}
\toprule
\textbf{Truth} & \textbf{Model} & \textbf{Key words} \\ 
\midrule
phishing & phishing & \textbf{The email:} "\hl{Someone can access your paypal account, so please confirm your identity to protect your account.} Start shopping faster by adding a payment method email customer paypal secure? Now check the account information that belongs to you! Why is my account access limited? Your account access has been limited for the following reason (s) we need to confirm some of your account information. Your case id, for this reason, is pp 009 536 987 252. We face a problem in the ratification of the real owner of the account and for this reason. We require your prompt attention and cooperation." \newline
\textbf{Note:} This phishing email tries to convey a sense of urgency and asks the email user to confirm their identity and provide account information. It also creates a sense of panic (loss) if the user does not act quickly. Our method successfully figures out the sentence representing the key message of the phishing email for a short and comprehensive explanation.\\
\midrule
phishing & phishing & \textbf{The email:} "Bulk attention! Your discover account will close soon! \hl{Dear member, we have faced some problems with your account, so please update the account.} If you do not update will be closed. To update your account, just confirm your information. (it only takes a minute). It's easy 1. Click the link below to open a secure browser window. 2. Confirm that you're the owner of the account, and then follow the instructions." \newline
\textbf{Note:} The message from the selected sentence obtained from our method exhibits cognitive triggers commonly associated with phishing attempts used in the phishing email. In particular, it implies a sense of urgency (concern) via "problems with your account while "dear member" aims to establish a connection with the recipient and imply that the message comes from a trusted source. The phrase "please update the account" creates a sense of familiarity and consistency.\\ 
\midrule
phishing & phishing & \textbf{The email:} "Your account limitation. \hl{We regret to inform you that access to your account has been temporarily limited.} This has been done due to several failed log in attempts. Case id ax 309 05 66 to restore your account please log in correctly. If you fail to log in correctly your account will be suspended for fraud prevention. You will be able to register again for paypal only after you authenticate your profile. We apologize for the inconvenience. This measure was taken for your protection. Paypal security team."\newline
\textbf{Note:} The selected sentence conveys a sense of urgency by stating that "access to your account has been temporarily limited." This urgency can trigger an emotional response and prompt the recipient to take quick action without thorough consideration. Furthermore, the message's phrasing, "we regret to inform you," and "access to your account has been temporarily limited," uses action-oriented language that urges the recipient to take immediate action to address the issue. In short, the model has successfully figured out the important sentence containing the cognitive principles in the phishing email.\\


\midrule
benign & phishing & \textbf{The email:} "Lower your mortgage payment! Bad credit, no problem! Lower your mortgage payment! Bad credit, no problem! Whether your credit is excellent or less than perfect, loanweb has a lender that can help you! Lowest rates on the web! Bad credit? Refinance to get cash! Lower monthly payments! Bad credit? Refinance to consolidate bills! \hl{Click here for lower mortgage payment!} Win a free bread maker! Win free bread makers, toaster ovens, cookware and more get a free subscription to cooking pleasures magazine get a free multi purpose grater test and keep free cooking products get free recipes from world famous chefsplus get a free 90 day membership in the cooking club of america. Home equity loans without perfect credit! Click here for a home equity loan! Free application! Home equity loans up to 125 potential tax deductible interest! Customized, competitive equity lines and loans. Apply now!"\newline
\textbf{Note:} 
In the used datasets, the ground truth label for this email is benign. However, our method predicts it as phishing due to the detection of potential phishing indicators, such as the use of enticing offers of "lowering mortgage payments" and the presence of "suspicious links". These elements may prompt users to provide personal information and bank account details via the phrase "Click here for lower mortgage payment!". Although the email's ground truth classification was incorrect, we deem the predicted label and the highlighted sentence from our model valuable in alerting users to potential phishing attempts.\\

\bottomrule
\end{tabular}
}
\vspace{-8mm}
\end{table}

\revise{Note that under a weakly supervised setting, when dealing with complex data (e.g., emails written from various writing styles and structures) and only the most important (top-1) selected sentences from emails are utilized, the observed improvement of our \ourapp~method from around 1.5\% to 3.5\% in the combined average performance of the Label-Accuracy and Cognitive-True-Positive measures, especially within the range over 99\% and approaching 100\%, signifies a substantial advancement}.

\vspace{-1mm}
\paragraph{\textbf{Qualitative Results}}

To demonstrate the effectiveness of our \ourapp~ method in solving the phishing attack localization problem, in Table \ref{tab:visualizationresult1}, we present various email samples alongside the most important (top-1) selected sentences extracted by our method as well as the predicted labels of the corresponding emails based on the selected sentences. Via these qualitative results, our method showcases \textit{its effectiveness in learning and figuring out the most important and phishing-relevant information (i.e., sentences) from phishing emails. That helps provide useful and concise interpretations for the phishing prediction}. Via these phishing email samples, it is evident that the commonly used cognitive triggers in the phishing emails related to scarcity (i.e., creates a sense of scarcity and urgency), authority (i.e., creates a sense of trust and legitimacy), and consistency (i.e., creates a sense of familiarity and consistency) principles, are also represented via the selected sentences. In addition, we also observe that the most important (top-1) sentence causing a specific email phishing tends to appear in some first positions of the email to get the users' attention.

\revise{In summary, the quantitative results in Table \ref{tab:my_label1} demonstrate the high quality of our \ourapp~ method for phishing attack localization, as evidenced by the Label-Accuracy and Cognitive-True-Positive measures. Additionally, the qualitative findings presented in Table \ref{tab:visualizationresult1} underscore the effectiveness of our \ourapp~ method in providing concise yet important explanations about phishing attacks. The most important (top-1) selected sentences from the emails offer valuable insights into the nature of the attacks.

In addition, in Table \ref{tab:visualizationresult1}, we also present an example where our \ourapp~ method incorrectly predicts the ground truth of the email. Specifically, the ground truth of this email is labeled as benign. However, our \ourapp~ method predicts it as phishing. Through our observations and the top-1 selected sentence obtained from our method, we believe that the model has identified some potential phishing attempts within the email, i.e., the use of enticing offers of "lowering mortgage payments" and the presence of "suspicious links". Therefore, despite the incorrect classification of the email’s ground truth, we consider the predicted label and the highlighted sentence from our model valuable in alerting users to possible phishing threats. Leveraging the predicted label and selected sentence, users can take proactive measures, such as contacting the loanweb directly using a verified phone number or email address, to verify the message's authenticity.}

\vspace{3mm}
\colorbox{gray!20}{
\begin{minipage}{0.9\columnwidth}
\textbf{In conclusion for RQ1}: The quantitative experimental results in Table \ref{tab:my_label1} on two main measures (i.e., Label-Accuracy and Cognitive-True-Positive) and the qualitative results in Table \ref{tab:visualizationresult1} show the superiority of our \ourapp~ method in achieving high performances for phishing attack localization on the real-world email datasets over the baselines. Our method demonstrates its effectiveness in predicting the vulnerability label (i.e., phishing or benign) of emails as well as in figuring out the most important and phishing-relevant information in the emails to give a useful and concise explanation about the vulnerability prediction for users.
\end{minipage}
}
\vspace{3mm}

Please refer to the appendix for the additional qualitative experiments (Section \ref{sec:aqresults}) of our \ourapp~ method on other phishing-related email samples.

\vspace{2mm}
\textbf{RQ2: \rqtwo}
\vspace{-0mm}

\paragraph{\textbf{Approach}}

We compare the performance of our \ourapp~ method with the baselines (i.e., \textbf{L2X} \cite{Chen2018-l2x}, \textbf{INVASE} \cite{INVASE2019}, \textbf{ICVH} \cite{nguyen2021-icvh}, \textbf{VIBI} \cite{Bang2021-vibi}, and \textbf{AIM} \cite{Vo2023-aim}) on the SCA measure where we aim to investigate if our \ourapp~ method consists of more important and commonly used cognitive triggers \revise{in the most important (top-1) selected sentences from corresponding emails than the baselines}.

\begin{table}[th]
    \centering{}
    \vspace{-2mm}
    \caption{The performance of our \ourapp~ method and the baselines for the SAC measure on the testing set (the
best results in \textbf{bold}).}
    \label{tab:my_label2}
    \vspace{-3mm}
    \resizebox{0.4\columnwidth}{!}{
    \begin{tabular}{c|c}
    \hline
    \textbf{Methods} & \textbf{SAC}  \\
    \hline
    INVASE \cite{INVASE2019} & 88.57\% \\
    \hline
    ICVH \cite{nguyen2021-icvh} &	83.88\% \\
    \hline
    L2X \cite{Chen2018-l2x} &	88.56\% \\
    \hline
    VIBI \cite{Bang2021-vibi} & 82.79\% \\
    \hline
    AIM \cite{Vo2023-aim} & 83.57\% \\
    \hline
    \ourapp~ (Ours) & \textbf{89.22\%} \\
    \hline
    \end{tabular}}
\vspace{-3mm}
\end{table}

\paragraph{\textbf{Results}}

The results in Table \ref{tab:my_label2} show that our \ourapp~ method obtains \textit{the best performance} on the SAC measure compared to the baselines. In particular, \revise{through the most important (top-1) selected sentences from corresponding emails,} our \ourapp~ method achieves higher performances than the baselines up to 6.5\% (when compared to the VIBI method). \revise{These results prove the effectiveness of our method not only in figuring out the most meaningful and crucial information causing the vulnerability of the email data (as shown in Table \ref{tab:my_label1}) but also in consisting of the more popular and important cognitive triggers used in phishing emails through the corresponding top-$1$ selected sentences compared to the baselines.}

\vspace{2mm}
\colorbox{gray!20}{
\begin{minipage}{0.9\columnwidth}
\textbf{In conclusion for RQ2}: The experimental results in Table \ref{tab:my_label2} demonstrate that our \ourapp~ method primarily consists of the commonly used and important cognitive principles in phishing emails compared to those of the baselines via the most important (top-1) selected sentences. That facilitates providing users with the most crucial and useful information for understanding to defend against phishing attacks.
\end{minipage}
}
\vspace{2mm}

\vspace{1mm}
\textbf{RQ3: \rqthree}

\vspace{0mm}
\paragraph{\textbf{Approach}}

We compare the performance of our proposed method to itself in the cases of using and without using our proposed information bottleneck theory additional training term mentioned in Eq. (\ref{eq:max_info-2}) and the data-distribution mechanism described in Eq. (\ref{eq:realx}). 

\vspace{1mm}
We aim to investigate if these additional terms successfully solve the potential limitations of the training principle mentioned in Eq. (\ref{eq:max_info}) for further improving the data representation learning that supports the process of learning and figuring out the important and true phishing-relevant information (i.e., sentences) in phishing-relevant emails. Note that in the case without using our introduced information bottleneck theory additional training and data-distribution mechanism terms, we denote the method as \ourapp-w/oIBT2D.

\paragraph{\textbf{Results}}

\begin{table}[th]
    \centering{}
    \vspace{-2mm}
    \caption{The performance of our method to itself in the cases of using and without using our information bottleneck theory additional training term and the data-distribution mechanism for the Label-Accuracy (Label-Acc) and Cognitive-True-Positive (Cognitive-TP) measures on the testing set (the higher
results in bold).}
    \label{tab:my_label3}
    \vspace{-3mm}
    \resizebox{0.9\columnwidth}{!}{
    \begin{tabular}{c|c|c|c}
    \hline
    \textbf{Methods} & \textbf{Label-Acc} & \textbf{Cognitive-TP} & \textbf{SAC} \\
    \hline
    \ourapp-w/oIBT2D & 98.92\% & 98.10\% & 88.61\% \\
    \hline
    \ourapp & \textbf{99.33\%}	& \textbf{98.95\%} & \textbf{89.22\%}\\
    \hline
    \end{tabular}}
\vspace{-1mm}
\end{table}

The results in Table \ref{tab:my_label3} demonstrate the effectiveness of our information bottleneck theory additional training term and the data-distribution mechanism in boosting the data representation learning that facilitates and supports the process of learning and figuring out the most important and true phishing-relevant information (i.e., sentences) in the corresponding emails. 

\vspace{1mm}
\revise{Although working within the context of only utilizing the top-1 selected sentences from thousands of diverse emails, particularly under the weakly supervised setting, these terms can still help improve the performance of \ourapp~ in all measures including Label-Accuracy, Cognitive-True-Positive, and SAC. The \ourapp~ method (with our information bottleneck theory additional training term and the data-distribution mechanism) approaches 99.5\% for Label-Accuracy and achieves around 99\% for Cognitive-True-Positive. These improvements are valuable. The results help contribute to the overall performance of our \ourapp~ method to obtain state-of-the-art performance on the Label-Accuracy, Cognitive-True-Positive, and SAC measures compared to the baselines as shown in Table \ref{tab:my_label1}.}

\vspace{2mm}
\colorbox{gray!20}{
\begin{minipage}{0.9\columnwidth}
\textbf{In conclusion for RQ3}: The experimental results in Table \ref{tab:my_label3} show the effectiveness of our bottleneck theory training term and the data-distribution mechanism in boosting the data representation learning process that is beneficial for not only the emails' vulnerability prediction but also for automatically figuring out the truly crucial and phishing-relevant information (via the most important one) in phishing emails compared to the case without using these terms.
\end{minipage}
}
\vspace{2mm}

\paragraph{\textbf{Training time}}

The training time of our model in the cases of using and without using our proposed information bottleneck theory training term and data-distribution mechanism is considerably short, costing approximately 36.47 seconds and 21.99 seconds, respectively, for each epoch (as mentioned in Section \ref{sec:modelc}, we ran our experiments on a 13th Gen Intel(R) Core(TM) i9-13900KF having 24 CPU Cores at 3.00 GHz with 32GB RAM, integrated Gigabyte RTX 4090 Gaming OC 24GB).

\subsection{Human evaluation}\label{sec:humane}

We additionally conduct a human evaluation to investigate the usefulness of our proposed \ourapp~ method in figuring out important and phishing-related information (e.g., sentences) in phishing emails to help provide comprehensive explanations about the phishing of the corresponding email data. In particular, we evaluate whether the most important (top-1) selected sentence in each phishing email by our proposed \ourapp~ method is perceived as convincing information for email users. To do that, we asked participants to evaluate the selected sentences \textit{of $10$ different phishing emails} (randomly chosen from the testing set) in terms of whether the sentence selected in each email is important to influence and persuade users to follow the instructions in the email.

\begin{figure}[h]
\begin{centering}
\includegraphics[width=0.95\columnwidth]{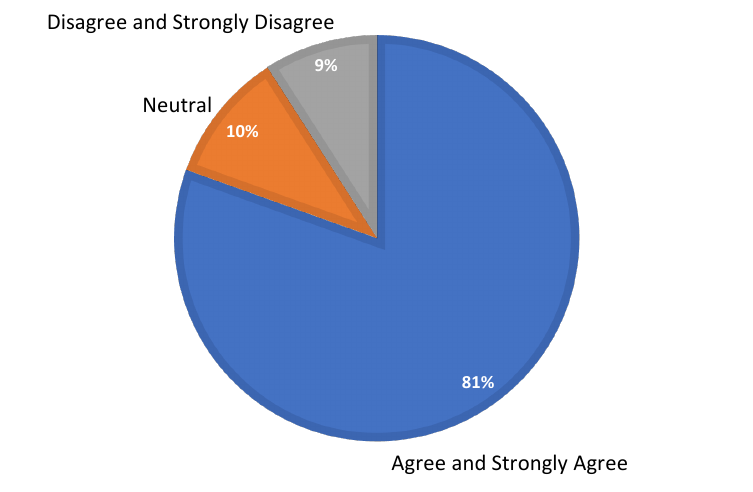}
\par\end{centering}
\vspace{-3mm}
\caption{Human evaluation on the importance of the top-$1$ selected information (i.e., a sentence) from each email (by our \ourapp~ method) in affecting and persuading users to follow the instructions from the email. We evaluate the selected sentences of 10 different phishing emails (randomly chosen from the testing set).  \label{fig:human-evaluation}}
\vspace{-3mm}
\end{figure}

There were \textit{$25$ university students and staff} participating in our survey. All participants responded that they have used emails for work and study, and they have both experienced and heard about phishing attacks. Based on the responses, in summary, as depicted in Figure \ref{fig:human-evaluation}, 81\% of participants selected either "Agree" (55\%) or "Strongly Agree" (26\%) when asked if they believe the selected sentences affect users' decision to follow the instructions in the survey phishing emails (note that in each phishing email, we use the top-1 selected sentence obtain from our \ourapp~ method). In contrast, 10\% of participants chose "Neutral", while 9\% chose "Disagree" or "Strongly Disagree". These human evaluations demonstrate the effectiveness of our method in successfully figuring out important phishing information (the attackers use it to deceitfully persuade the target users). The selected information in each phishing helps provide a useful and comprehensive explanation of the vulnerability prediction of the email, especially for the phishing one, to users. In essence, this approach enhances the explainability (transparency) of phishing detection, addressing and improving the overall clarity of the prediction process.

Note that, in this human evaluation, we implemented careful study design protocols to minimize potential priming. To ensure objectivity in the results, no information was provided about the source of the selected sentences.

\section{Threats to validity}
\label{sec:threats}

\paragraph{\textbf{Construct validity}}
Key construct validity threats are if the assessments of our proposed method and baselines demonstrate the ability for phishing attack localization. In our paper, we study an important problem of phishing attack localization where we not only aim to automatically learn and figure out the most important and phishing-relevant information (e.g., sentences) in each phishing email but also can detect the vulnerability $Y$ (i.e., phishing or benign) of the corresponding email based on the crucial selected information. The selected phishing-relevant information (e.g., sentences) helps provide useful and concise explanations about the vulnerability of the phishing email data. To evaluate the performance of our method and baselines, we use three main measures including Label-Accuracy, Cognitive-True-Positive, and SAC.

\vspace{-1mm}
\paragraph{\textbf{Internal validity}}
Key internal validity threats are relevant to the choice of hyper-parameter settings (i.e., optimizer, learning rate, number of layers in deep neural networks, etc.). It is worth noting that finding a set of optimal hyperparameter settings of deep neural networks is expensive due to a large number of trainable parameters. To train our method, we only use the common and default values of hyper-parameters such as using Adam optimizer and the learning rate equals $10^{-3}$. We also report the hyperparameter settings in our released reproducible source code samples to support future replication studies.

\vspace{-1mm}
\paragraph{\textbf{External validity}}
Key external validity threats include whether our proposed \ourapp~ method can generalize well to different types of phishing email datasets. We mitigated this problem by conducting our experiments on seven real-world diverse phishing-relevant email datasets including IWSPA-AP, Nazario Phishing Corpus, Miller Smiles Phishing Email, Phish Bowl Cornell University, Fraud emails, Cambridge, and Enron Emails.
\section{Future work}\label{sec:future-work}
\revise{In this study, our paper focuses primarily on addressing the phishing attack localization problem in email phishing, the most common form of phishing. It is worth noting that phishing attacks can also occur through webpages. Several methods have been proposed to detect and infer phishing intention based on webpage appearances (e.g., \cite{VisualPhishNet2020, Liuphishingintention2022}). We believe the principles underlying our \ourapp~ framework can also be extended to detect and explain webpage phishing attacks. The operational nature of our AI2TALE approach within (i) a unified framework, (ii) working directly with webpage data without requiring additional steps to gain extra information (e.g., \note{references to the ground truth of phishing information}) beyond the data and its vulnerability label (i.e., phishing or benign), as well as (iii) the way our AI2TALE model can be trained without requiring the ground truth of phishing information in the data can be considered as some of the advantages of our framework compared to the relevant methods (e.g., \cite{VisualPhishNet2020, Liuphishingintention2022}). Investigating the application of our \ourapp~ framework to explain webpage phishing attacks could be a focus of our future studies.}
\section{Related background}

In this section, we briefly present the main related background used in our proposed \ourapp~ method.

\subsection{\textbf{Mutual information}}

Mutual information (MI) is used to measure the dependence
between two random variables \cite{TheoryofC, EofIT2006}. It captures how much the knowledge of one random variable reduces the uncertainty of the other. In particular, MI quantifies the amount of information obtained about one random variable by observing the other random variable. For instance, suppose variable A signifies the outcome of rolling a standard 6-sided die, and variable B represents whether the roll yields an even (0 for even, 1 for odd) result. It is evident that information from B offers insights into the value of A, and vice versa. In essence, these random variables exhibit mutual information.

Assume that we have two random variables $X$ and $Y$ drawn from the joint distribution $p(x,y)$ with two corresponding marginal distributions $p(x)$ and $p(y)$. The mutual information between $X$ and $Y$ denoted by $I(X,Y)$ is the relative entropy between the joint distribution $p(x,y)$ and the product distribution $p(x)p(y)$, and is defined as follows:

\vspace{-1mm}

\begin{align*}
I(X,Y) & =\int p(x,y)\log\frac{p(x,y)}{p(x)p(y)}dxdy\\
 & =D_{KL}(p(x,y)||p(x)p(y))
\end{align*}

\vspace{1mm}
where $D_{KL}(p(x,y)||p(x)p(y))$ is the Kullback-Leibler divergence measuring the relative entropy (i..e, the difference in information) represented by two distributions, i.e., the product of marginal distributions $p(x)p(y)$ of $X$ and $Y$ from their joint distribution $p(x,y)$.

\subsection{\textbf{Information bottleneck theory}}

Here, we consider the supervised learning context where we want to predict corresponding outputs (e.g., labels) $\left\{ \by_{i}\right\} _{i=1}^{n}$ of given inputs $\left\{ \bx_{i}\right\} _{i=1}^{n}$. A deep learning network (DNN) will learn latent representations (i.e., latent features in the latent space that contain useful information to describe the data) $\left\{ \tilde{\bx}_{i}\right\} _{i=1}^{n}$ of the corresponding input data samples $\left\{ \bx_{i}\right\} _{i=1}^{n}$ in terms of enabling good predictions and generalizations.

\begin{figure}[h]
\vspace{-1mm}
\begin{centering}
\includegraphics[width=0.5\columnwidth]{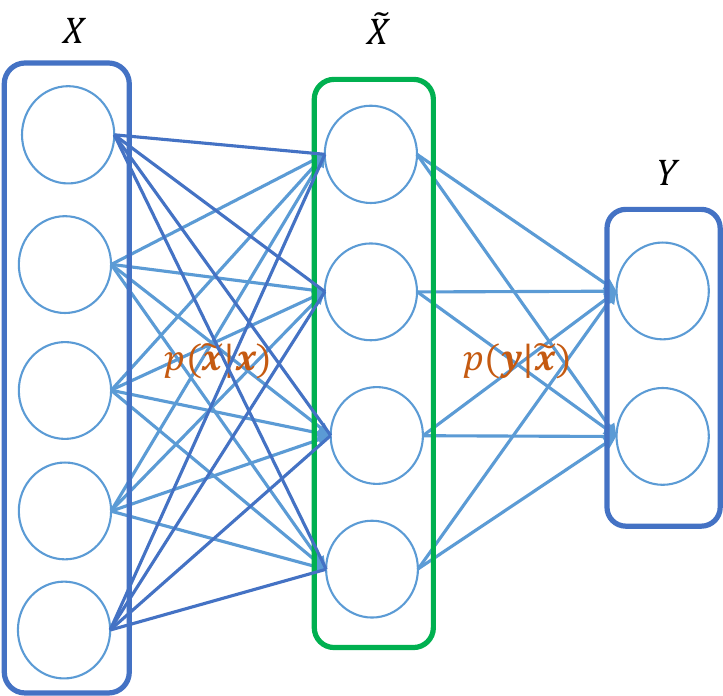}
\par\end{centering}
\vspace{-1mm}
\caption{An architecture of a simple deep neural network in a supervised learning context for the classification problem. \label{fig:bottneck-theory}}
\vspace{-1mm}
\end{figure}

Assume that the whole hidden layer in Figure \ref{fig:bottneck-theory} is denoted by a random variable $\tilde{X}$ while the input and output layers are denoted by random variables $X$ and $Y$ respectively. We can describe this hidden layer by two conditional distributions: the encoder $p(\tilde{\bx}|\bx)$ and the decoder $p(\by|\tilde{\bx})$. This transformation process preserves the information of the input layer $X$ without considering which individual neurons within the hidden layer $\tilde{X}$ encode which features (i.e., neurons) of $X$. An optimal encoder process of the mutual information between $X$ and the desired output $Y$ denoted by $I(X,Y)$ can create the most compact encoding (i.e., minimally sufficient statistic) $\tilde{X}$ of the input data $X$ while $\tilde{X}$ still has enough information (i.e., $\tilde{X}$ can capture the important features of $X$ as well as remove the unnecessary parts of $X$ that do not make any contributions to the prediction of $Y$) to predict
$Y$ as accurately as possible.

An information bottleneck \citep{tishby2000information,tishby2015deep}
is proposed to be a computational framework that aims to find the most compact encoding $\tilde{X}$ of the input data $X$. In particular, it is the optimal trade-off between the compression $\tilde{X}$ and the prediction of the desired output $Y$ as described in the following optimization problem:

\begin{equation*}
\min_{p(\tilde{\bx}|\bx),p(\by|\tilde{\bx}),p(\tilde{\bx})}\left\{ I(X,\tilde{X})-\beta I(\tilde{X},Y)\right\} 
\end{equation*}

where $\beta$ specifies the amount of relevant information captured by the encoding process (i.e., the representations $\tilde{X}$ and $I(\tilde{X}, Y)$).
\section{Conclusion}\label{sec:conclusion}

In this paper, we have successfully proposed an innovative deep learning-based method derived from an information-theoretic perspective and information bottleneck theory for solving the phishing attack localization problem where automated AI-based techniques have not yet been well studied. Our \ourapp~ method can work effectively in the weakly supervised setting providing an important practical solution to solve the problem in terms of not only accurately predicting the vulnerability of the email data but also automatically learning and figuring out the most important and phishing-relevant information in each phishing email. The selected information provides useful and concise explanations about the vulnerability of the corresponding phishing email data. In addition, we also introduce appropriate measures for phishing attack localization. The rigorous and comprehensive experiments on seven real-world diverse email datasets show the superiority of our proposed \ourapp~ method over the state-of-the-art baselines.

\bibliographystyle{ACM-Reference-Format}
\bibliography{reference}

\section{Appendix}\label{sec:appendix}

\subsection{Additional quantitative results}\label{sec:aqresults}

In our study, as presented in RQ1, via the Label-Accuracy metric, we aim to measure whether the most important (top-1) selected sentences obtained from each model can accurately predict the associated emails' vulnerability label (i.e., phishing or benign). The underlying intuition is that the most pivotal sentences significantly influence the vulnerability of emails, particularly phishing ones. Specifically, in the experiments, for each email, we focus on evaluating the most important (top-1) selected sentence (i.e., aiming to provide the most highly qualified and concise explanation of the vulnerability of email data to users) from each model to determine its efficacy in predicting the email's vulnerability label, without considering all of its sentences. The higher the value of the Label-Accuracy measure, the better the model’s performance in selecting the most crucial and label-relevant information from the data.

To examine our experiments further on this aspect, we calculated the F1-score for our \ourapp~ method and baselines. Specifically, we computed the F1-score in two scenarios: (i) exclusively on the testing phishing emails, and (ii) on both the testing phishing and benign emails. The results for our \ourapp~ method and the baselines (i.e., AIM, VIBI, L2X, ICVH, and INVASE) in scenarios (i) and (ii) are shown in Table \ref{tab:my_label1_adq}. 

\begin{table}[th]
    \centering{}
    \vspace{1mm}
    \caption{The performance of our \ourapp~ method and the baselines for the F1-score measure on the testing set in two scenarios (i) and (ii) (the
best results in \textbf{bold}).}
    \label{tab:my_label1_adq}
    \vspace{-3mm}
    \resizebox{0.68\columnwidth}{!}{
    \begin{tabular}{c|c|c}
    \hline
    \textbf{Methods} & \textbf{F1-score (i)} & \textbf{F1-score (ii)}  \\
    \hline
    INVASE \cite{INVASE2019} & 98.313\% & 98.299\% \\
    \hline
    ICVH \cite{nguyen2021-icvh} &	96.732\%	& 96.725\% \\
    \hline
    L2X \cite{Chen2018-l2x} &	98.261\%	& 98.249\% \\
    \hline
    VIBI \cite{Bang2021-vibi} & 96.626\% & 96.649\% \\
    \hline
    AIM \cite{Vo2023-aim} & 98.406\% & 98.399\% \\
    \hline
    \ourapp~ (Ours) & \textbf{99.324\%}	& \textbf{99.325\%} \\
    \hline
    \end{tabular}}
    \vspace{-3mm}
\end{table}

In summary, the F1-score results for our \ourapp~ method and baselines closely align with those obtained using the accuracy measure (i.e., Label-Accuracy). Notably, our \ourapp~ method exhibits consistent effectiveness and advancement, considerably surpassing the baselines in both scenarios (i) and (ii).

\vspace{1mm}
\subsection{Additional qualitative results}\label{sec:aqresults}

\begin{table}[!h]
\caption{The ground truth and the predicted label based on the most important (top-1) selected sentence (highlighted in yellow) of each email from our \ourapp~ method.}
\vspace{-1mm}
\label{tab:visualizationresult2}
\vspace*{0.0em}
\centering
\resizebox{1.0\linewidth}{!}{%
\begin{tabular}{p{1cm}|p{1cm}|p{9.9cm}}
\toprule
\textbf{Truth} & \textbf{Model} & \textbf{Key words}
\\
\midrule
phishing & phishing & \textbf{The email:} "Multiple number of incorrect login attempts on your halifax account. Dear halifax online customer. \hl{Your account is suspended due to multiple number of incorrect login attempts.} For your protection, we've suspended your account. To restore your account please log in correctly. If you fail to log in correctly your account will be suspended for fraud prevention. Sincerely, halifax online helpdesk copyright halifax bank a little extra help."  
\newline
\textbf{Note that:} The email tries to convey a sense of urgency (or loss) and asks the user to log in to their account to avoid suspension or protect their account from any unknown accessing attempts. This crucial message is represented in the selected sentence. That shows the effectiveness of the proposed method in figuring out the most relevant phishing sentence in the phishing email supporting a comprehensive explanation.
\\
\midrule
phishing & phishing & \textbf{The email:} "Temporarily suspended. Dear customer, customer advice, please address the following issues. The details that you have entered have not been recognized. \hl{For your security, your online service has been temporarily locked.} No further attempts will be accepted. If you provide us with the following details, you should be able to access the service in just a few minutes click here to get started legal info privacy security 2005 2010." 
\newline
\textbf{Note:} This phishing email tries to convey (i) the action taken is intended to protect the recipient's well-being and (ii) a sense of urgency, encouraging the recipient to address the issue promptly to regain access to their online service. Our method successfully figures out the sentence representing the key message of the phishing email for a short and comprehensive explanation.
\\
\midrule
phishing & phishing & \textbf{The email:} "Dear paypal customer. \hl{Dear paypal customer, this is an official notification from paypal that the service listed below will be deactivated and deleted if not renewed immediately.} Previous notifications have been sent to billing contact assigned to this account. As the primary contact, you must renew the service listed below, or it will be deactivated and deleted. Click to renew your paypal account now service paypal security department. Expiration may 14, 2010, at paypal we are dedicated to providing you with exceptional service and to ensuring your trust. If you have any questions regarding our services, please check the website or call our customer service. Thank you, sincerely, paypal security department paypal 's services terms and conditions apply. The information on this page is presented subject to our legal page and any other terms and conditions that paypal may impose from time to time. It is subject to change without notification. Microsoft and the microsoft internet explorer are registered trademarks or trade works of microsoft corporation in the united states and for other countries."  
\newline
\textbf{Note that:} The selected sentence employs urgent (fear and concern) language by stating that the service will be "deactivated and deleted if not renewed immediately." This urgency can create panic and pressure the recipient to act quickly without considering the legitimacy of the message. Furthermore, it also attempts to establish credibility and authority via "official notification from paypal" often used by attackers. The message from the selected sentence starts with the generic address "dear paypal customer" twice. While legitimate communications would typically use the recipient's actual name, phishing emails often lack personalization and use general greetings. In general, the selected sentence consists of almost all key messages from the email. That shows the effectiveness of the proposed method in figuring out the most important phishing-relevant information (sentence) from the email.
\\

\bottomrule
\end{tabular}
}
\vspace{-1mm}
\end{table}

In Table \ref{tab:visualizationresult2}, we show qualitative experimental results of our \ourapp~ method on some further phishing email samples. Via these qualitative results, our method again demonstrates \textit{its effectiveness in learning and figuring out the most important and phishing-relevant information (i.e., sentences) from phishing emails for providing useful and concise interpretations corresponding to the predicted phishing labels}. Through these phishing email samples, it is also shown that the most important (top-1) selected sentences from the model primarily consist of the commonly used cognitive triggers in the phishing emails related to scarcity (i.e., \textit{creates a sense of scarcity and urgency}), authority (i.e., \textit{creates a sense of trust and legitimacy}), and consistency (i.e., \textit{creates a sense of familiarity and consistency}) principles. From these samples, we also observe that the most important sentence causing a specific email phishing tends to appear in the initial positions of the email to capture the users' attention.
\end{document}